\documentclass[%
 reprint,
superscriptaddress,
 amsmath,amssymb,
 aps,
 prl,
floatfix,
]{revtex4-2}
\usepackage{cancel}
\usepackage[caption=false]{subfig}
\usepackage{graphicx}
\usepackage{dcolumn}
\usepackage{bm}
\usepackage[hidelinks,breaklinks]{hyperref}
\usepackage{xcolor}
\usepackage[normalem]{ulem}
\definecolor{ra}{rgb}{0.8, 0.0, 0.0}

\usepackage{multirow}
\usepackage{pdfpages}
\newcommand{\pairshift}{0pt}
\newcommand{\pair}[1]{\multirow{2}{*}[\pairshift]{#1}}
\makeatletter
\AtBeginDocument{\let\LS@rot\@undefined}
\makeatother

\newcommand{\avg}[1]{\left\langle #1\right\rangle}
\begin{document}

\preprint{APS/123-QED}

\title{Exact and Near-Exact Inverse Renormalization Group with Constrained Monte Carlo}

\author{Dimitrios Bachtis}
\email{db2091@cam.ac.uk}
\affiliation{Department of Applied Mathematics and Theoretical Physics, University of Cambridge, Wilberforce Road, Cambridge CB3 0WA, United Kingdom}

\include{ms.bib}

\date{November 5, 2026}

\begin{abstract}

We establish conditions for exactness of the inverse renormalization group  via constrained Markov processes, a class of Monte Carlo simulations which sample the degrees of freedom within a block on a fine lattice under the constraint that they remain consistent with the corresponding degree of freedom on the coarse lattice according to a forward renormalization group transformation. We show analytically, for an arbitrary transformation, that constrained Markov processes correspond to an exact inversion of the renormalization group at the fixed point action, and to a near-exact inversion in the vicinity of the fixed point. We then optimize renormalization group transformations directly on inversely renormalized configurations to effectively map the action onto itself and ensure computational stability in the generation of configurations for high-dimensional systems. Starting from a set of uncorrelated configurations for the three-dimensional $\phi^{4}$ quantum field theory at lattice volume $V=16^{3}$, we invert the renormalization group with constrained Monte Carlo to construct a set of uncorrelated, near-equilibrium configurations for $V=128^{3}$ and extract two critical exponents of the system. The inversion exhibits no critical slowing down, with dynamic critical exponent $z=0.03(2)$.

\end{abstract}

\maketitle

\paragraph{\label{sec:level1}Introduction.---}

Inverse renormalization group methods construct large-volume configurations from small ones, hence overcoming the critical slowing down effect. Here we establish when this inversion is exact. Inverse renormalization group methods were first validated on Ising models by Ron-Swendsen-Brandt~\citep{PhysRevLett.89.275701}. Related multiscale methods, in which coarse configurations are refined and then rethermalized, were proposed to accelerate the equilibration of $\mathrm{SU}(N)$ lattice gauge theories~\citep{PhysRevD.92.114516}. Inverse renormalization group transformations were subsequently formulated in quantum field theory with convolutional neural networks~\citep{PhysRevLett.128.081603}, and have been applied to statistical mechanical models, disordered systems, and low-dimensional quantum field theories~\citep{PhysRevB.99.075113,Shiina2021,3njc-5tlx,PhysRevX.13.041038,PhysRevB.110.L140202}.

To our knowledge, there exist two studies to have pursued a direct inversion of a forward renormalization group transformation on high dimensional systems with nontrivial renormalization group operator content. These are the original work of Ron-Swendsen-Brandt on the 3D Ising model~\citep{PhysRevLett.89.275701}, and the study of the 3D Edwards-Anderson model by the author~\citep{PhysRevB.110.L140202}. Of the aforementioned studies only Ron-Swendsen-Brandt pursue what can be interpreted as a genuine, and not approximate, inversion of a renormalization group transformation by relying on the implementation of a ``compatible Monte Carlo" method to construct configurations of the inversely renormalized system which remain, under the application of a forward renormalization group transformation, consistent with the original system. 

\begin{figure}[t]
\includegraphics[width=6cm]{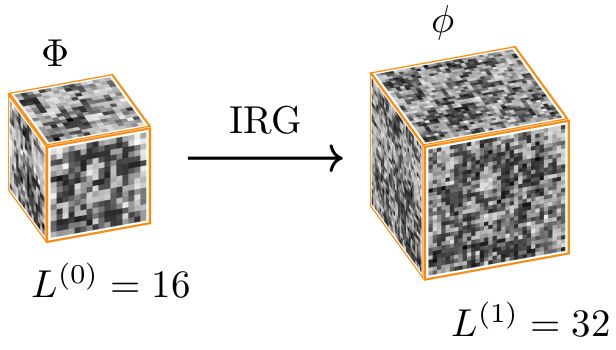}
\caption{\label{fig:fig1} Inversion of the renormalization group via constrained Monte Carlo simulations for the three-dimensional $\phi^{4}$ theory at the numerically calculated critical point. A coarse configuration $\Phi$ is provided as input to obtain a fine configuration $\phi$.  }
\end{figure}

Nevertheless, the Ron-Swendsen-Brandt implementation of the inverse renormalization group has certain limitations. It requires the simulation of a complicated fixed point Hamiltonian with $17$ coupling constants, instead of the original
Hamiltonian of the physical theory, in order to establish computational stability in high-dimensional systems. The direct simulation of a renormalized Hamiltonian was criticized as a limitation~\citep{PhysRevLett.42.859} in the Monte Carlo renormalization group proposal of Ma~\citep{PhysRevLett.37.461}, and it is currently manifest in the Ron-Swendsen-Brandt implementation of the inverse renormalization group. 

In this Letter we establish conditions for exactness of the inverse renormalization group via constrained Monte Carlo simulations of the original lattice action. We investigate the structure of the conditional probability distribution of a coarse degree of freedom given a  fine degree of freedom in the inversely renormalized system and its reverse definition. Specifically, we explore if we can analytically show that constrained Monte Carlo methods correspond to an exact inversion of the renormalization group for the case of the fixed point action and to a near-exact inversion in the vicinity of the fixed point.

We then discuss benefits related to the optimization of real-space renormalization group transformations directly on the space of the inversely renormalized configurations. We explore if we can construct an optimized transformation that effectively maps the action onto itself~\citep{PhysRevLett.52.2321}, hence avoiding the need to simulate a fixed point action with multiple terms in order to establish stability of the inverse renormalization group in high-dimensional systems. Starting from a set of uncorrelated configurations at lattice volume $V=16^{3}$ we investigate if we can invert the renormalization group with constrained Monte Carlo to obtain a set of uncorrelated and near-equilibrium configurations for $V=128^{3}$, and if we can extract two critical exponents.

\paragraph{\label{sec:level3}Exactness of the inverse renormalization group.---}
We consider the three-dimensional $\phi^4$ lattice field theory with periodic boundary conditions described by the lattice action:
\begin{equation}
S(\phi) = -2\kappa\sum_{x}\sum_{\mu=1}^{3}\phi_x\phi_{x+\hat\mu}
          + \sum_x\Big[\phi_x^2+\lambda\big(\phi_x^2-1\big)^2\Big],
\label{eq:action}
\end{equation}
with volume $V=L^3$, where $L$ is the linear dimension.
We consider the case of $\lambda=1.1$ for which the critical point has been numerically calculated as $2\kappa_c=0.3750966(4)$~\citep{Mhasen}. We define a coarse configuration as $\Phi$ and a fine configuration as $\phi$ and the corresponding lattice sites as $X$ and $x$. The coarse configuration $\Phi$ can be a renormalized configuration obtained by applying a forward renormalization group transformation $p(\Phi | \phi)$ on the fine configuration $\phi$. 

We define the conditional probability distributions as:
\begin{equation}
p(\Phi|\phi)=\frac{p(\phi,\Phi)}{p(\phi)}, \ p(\phi|\Phi)=\frac{p(\phi,\Phi)}{p(\Phi)},
\end{equation}
where $p(\phi)$, $p(\Phi)$ are the marginal probability distributions obtained by integrating out $\Phi$ or $\phi$ respectively, and $p(\phi,\Phi)$ is the joint distribution. We consider the aforementioned distributions to represent the distribution of the fixed point action $S'$ of a given transformation.

In constrained Monte Carlo we start from a sampled coarse configuration $\Phi$ that remains fixed and we are seeking a Markov process that will complete the fine degrees of freedom $\phi_{x, x \in B}$ within a block $B_{X}$ on the fine system so that application of a forward renormalization group transformation $p(\Phi | \phi)$ on $B_{X}$ will reproduce the corresponding renormalized coarse degree of freedom $\Phi_{X}$. We consider the reverse definition of the conditional probability distribution of the fine field $\phi$ given the coarse degree of freedom $\Phi$:
\begin{equation}
p(\phi |\Phi)=\frac{e^{-S(\phi)}\,p(\Phi|\phi)}{Z_{\phi}\,p(\Phi)}.
\label{eq:posterior}
\end{equation}

In a Monte Carlo simulation the resulting probability distribution of the fine configurations, denoted here as $q(\phi)$, is obtained constructively via $p(\phi |\Phi)$ based on the numerically sampled distribution of the coarse degrees of freedom $p_{\rm MC}(\Phi)$ which we utilize as the input distribution and from which we initiate the constrained Monte Carlo approach as:
\begin{align}
q(\phi) &= \int\!\mathcal D\Phi\; p(\phi|\Phi)\, p_{\rm MC}(\Phi) \nonumber\\
        &= \int\!\mathcal D\Phi\; p(\phi)\, p(\Phi|\phi)\, \frac{p_{\rm MC}(\Phi)}{p(\Phi)} \nonumber\\
        &= \frac{e^{-S(\phi)}}{Z_\phi} \int\!\mathcal D\Phi\; p(\Phi|\phi)\, \frac{p_{\rm MC}(\Phi)}{p(\Phi)} .
\label{eq:q}
\end{align}

The resulting probability distribution can be equivalently expressed as $q(\phi)=p(\phi)\,w(\phi)$, where
$w(\phi)=\int\!\mathcal D\Phi\,p(\Phi|\phi)\,p_{\rm MC}(\Phi)/p(\Phi)$. At its fixed point action the considered
forward renormalization group transformation leaves the action invariant $S'=S$, where $S'$ denotes the fixed point action. If the input distribution of the
coarse degrees of freedom $p_{\rm MC}(\Phi)$ is the probability distribution of the fixed point action $p(\Phi)$ so that
$p_{\rm MC}(\Phi)=p(\Phi)$, then $w=1$, since $\int\!\mathcal D\Phi\,p(\Phi|\phi)=1$. The inversion of the
renormalization group is therefore exact and $q(\phi)\equiv p(\phi)$. 

If the input distribution is not that of the
fixed point action, namely $p_{\rm MC}(\Phi)\neq p(\Phi)$, then constrained Monte Carlo corresponds to a near-exact inversion in the vicinity of the fixed point. Specifically, $w$ depends on the fine degrees of freedom $\phi$ only through the block
variables on which the forward transformation $p(\Phi|\phi)$ is applied. For fixed block variables the sampling of fine degrees of freedom is therefore exact. Any deviation of $q(\phi)$ from equilibrium resides solely in the
distribution of the block variables, originating from the discrepancy between $p_{\rm MC}(\Phi)$ and $p(\Phi)$.

In more detail, since the inversely renormalized configurations, distributed as
$q(\phi)=\int\mathcal{D}\Phi\, p(\phi|\Phi)\, p_{\rm MC}(\Phi)$, and the equilibrium configurations,
distributed as $p(\phi)$, share the same conditional distribution $p(\phi|\Phi)$, integrating out $\Phi$
cannot increase their relative entropy,
\begin{equation}
D_{\rm KL}\big(q(\phi)\,\|\,p(\phi)\big) \le D_{\rm KL}\big(p_{\rm MC}(\Phi)\,\|\,p(\Phi)\big),
\end{equation}
so the deviation of the inversely renormalized configurations from equilibrium is bounded by the
deviation of $p_{\rm MC}(\Phi)$ from $p(\Phi)$, which vanishes at the fixed point.

\paragraph{\label{sec:level5}Optimization of the inverse transformation.---} We implement constrained Monte Carlo to invert the renormalization group transformation $p(\Phi|\phi)$ by starting from a coarse configuration $\Phi$ sampled with conventional Monte Carlo methods for lattice size $L$. We then initialize a fine system of lattice size $2L$, by approximately determining the values of each block~\footnote{See the Supplemental Material}. At a lattice site $i$ a new degree of freedom is proposed as  $\phi_{i}'=\phi_i+\delta$ with
$\delta$ uniform in $(-1,1)$ which we accept with probability:  
\begin{equation}
A=\min\ \Bigg(1,\ e^{-\Delta S}\,\frac{p(\Phi|\phi')}{p(\Phi|\phi)}\Bigg).
\end{equation}
\begin{figure}[t]
\includegraphics[width=8.2cm]{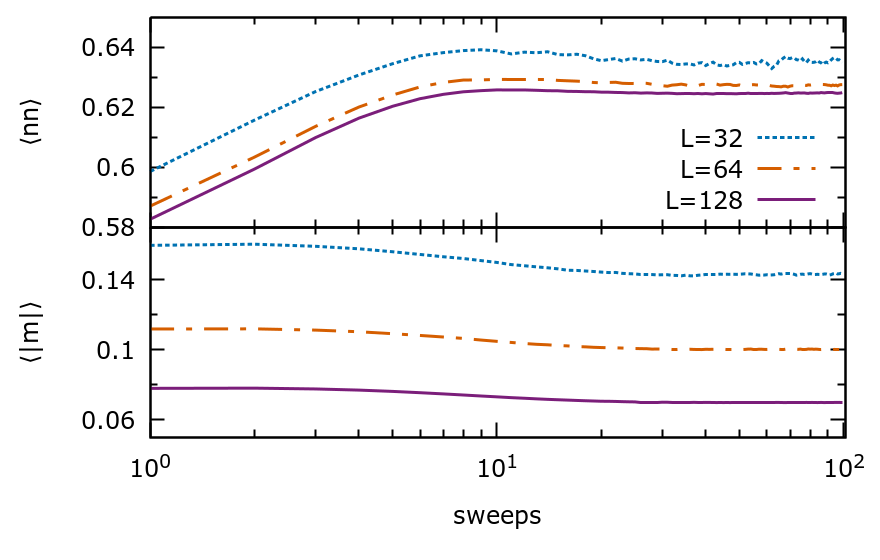}
\caption{\label{fig:fig2} Equilibration of the constrained Monte Carlo simulation for the inversion of the renormalization group $p(\phi | \Phi)$ for lattice sizes $32$, $64$, and $128$. The figures depict the nearest-neighbor operator $\langle nn \rangle$ and the absolute value of the magnetization $\langle|m|\rangle$ versus the number of sweeps.  }
\end{figure}
\begin{figure}[t]
\includegraphics[width=8.2cm]{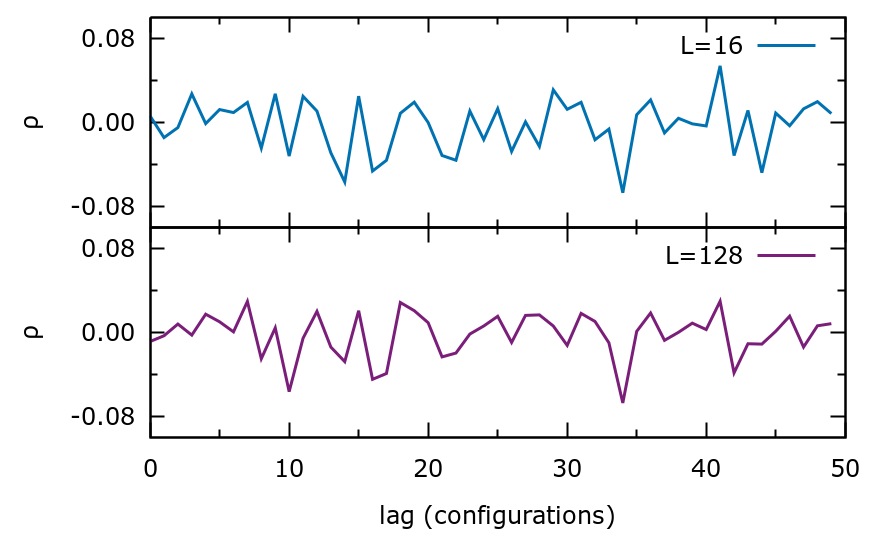}
\caption{\label{fig:fig3} Autocorrelation of the initial set of uncorrelated configurations at $L=16$  (top) and the $L=128$ configurations obtained after three consecutive inversions of the renormalization group (bottom). The autocorrelation is calculated based on the nearest-neighbor operator. }
\end{figure}
\begin{figure}[t]
\includegraphics[width=8.2cm]{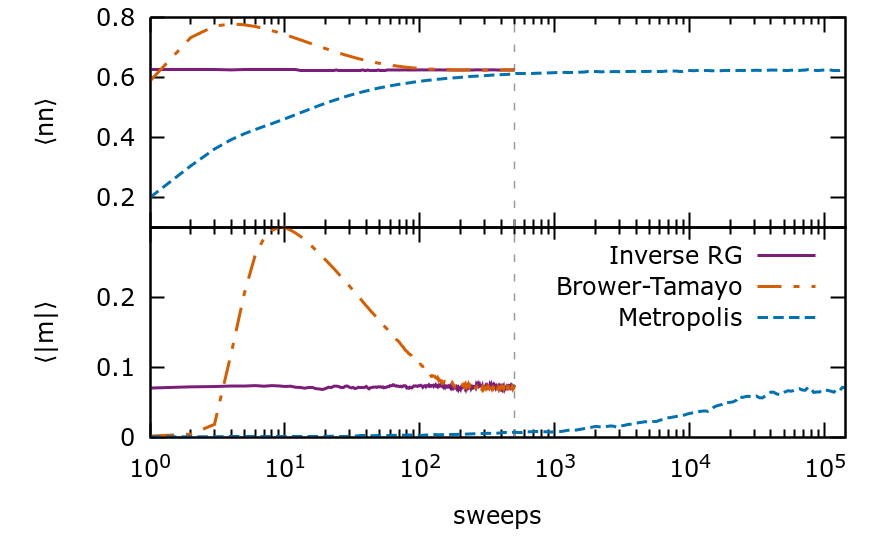}
\caption{\label{fig:fig4} Evolution of $500$ configurations obtained from the inverse renormalization group with the Brower-Tamayo algorithm with Swendsen-Wang updates. Comparison against a set of $500$ randomly initialized chains using Brower-Tamayo and, separately, $50$ Metropolis chains. The vertical dashed line corresponds to $500$ sweeps.}
\end{figure}

Here we consider a rescaling factor of $b=2$ for the inverse renormalization group and a transformation which combines elements from both linear and nonlinear versions. Specifically, we define the conditional probability distribution of coarse degrees of freedom $\Phi$ given the fine degrees of freedom $\phi$ as:
\begin{equation}
p(\Phi|\phi)=\prod_X\frac{1}{\sqrt{2\pi\sigma^2}}\exp\Big[-\frac{\big(\Phi_X-f(\widetilde{\Phi}_X)\big)^2}{2\sigma^2}\Big],
\label{eq:kernel}
\end{equation}
\begin{equation}
\widetilde{\Phi}_X=\sum_{ijk}K_{ijk}\,\phi_{ijk},\
f(\widetilde{\Phi})=\mathrm{sgn}(\widetilde{\Phi})\big(|\widetilde{\Phi}|^{\alpha}+c\,|\widetilde{\Phi}|^{3\alpha}\big),
\label{eq:smear}
\end{equation}
where  $\widetilde{\Phi}_X$ is a smeared field, obtained via a weighted sum of the fine degrees of freedom $\phi$ using kernel coefficients $K_{ijk}$. The considered kernel has size $6^{3}$ with its $2^{3}$ center positioned exactly at the corresponding block $B$ of size $b^{3}=2^{3}$ in the inversely renormalized fine system. The kernel has cubic symmetry, and the transformation respects the $\mathbb Z_2$ reflection symmetry. 

The kernel can be interpreted as an application of one strided convolution, utilized before in inverse renormalization group methods of lattice quantum field theories~\citep{PhysRevLett.128.081603}, and the function $f$ resembles the majority rule for the system with a dependence on tunable parameters $\alpha$ and $c$. The transformation comprises $13$ variational parameters  $\theta=(K_1,\dots,K_{10},\alpha,c,\sigma)$,  which are optimized with a gradient-based approach. Since the $6^{3}$ kernel consists of the block $B$ and its $26$ neighboring blocks, every proposed fine degree of freedom $\phi_{i}$ depends on the $27$ coarse degrees of freedom $\Phi$ of these blocks.

Constrained Monte Carlo for the inversion of the renormalization group can be equivalently expressed as a Metropolis Markov chain for the effective action:
\begin{equation}
S_{\rm eff}(\phi)=S(\phi)+\sum_X\frac{\big(\Phi_X-f(\widetilde{\Phi}_{X})\big)^2}{2\sigma^2},
\end{equation}
which constrains every block $B_{X}$ of the fine system to its corresponding coarse degree of freedom $\Phi_{X}$ based on a stochastic constraint. For $\sigma\to0$ the transformation becomes deterministic, $\Phi_X=f(\widetilde{\Phi}_X)$, and the constraint is
exact. For $\sigma>0$ the constraint is satisfied in distribution, with $\sigma$ a parameter of the transformation that
is optimized. In the Supplemental Material~\citep{Note1} we provide the values of the transformation parameters to allow reproducibility, and further discuss and clarify mathematical definitions and simulation parameters used throughout the manuscript.

\begin{table}[h]
\centering
\begin{tabular}{|c|c|c|c|c|}
\hline
Sweeps & $O$ & IRG-BT & IRG-M  & Rand-BT  \\
\hline
\pair{$0$}   & $\langle nn\rangle$ & $0.6252(2)$ & $0.6252(2)$ & -  \\ \cline{2-5}
             & $\langle |m|\rangle$ & $0.0708(15)$ & $0.0708(15)$ & - \\ \hline
\pair{$20$}  & $\langle nn\rangle$ &  $0.6243(2)$ & $0.6245(2)$ & $0.6957(2)$   \\ \cline{2-5}
             & $\langle |m|\rangle$ & $0.0697(16)$& $0.0707(15)$ & $0.2533(3)$  \\ \hline
\pair{$500$} & $\langle nn\rangle$ & $0.6245(2)$ & $0.6245(2)$ & $0.6244(2)$  \\ \cline{2-5}
             & $\langle |m|\rangle$ & $0.0715(16)$ & $0.0710(16)$ & $0.0707(16)$   \\ \hline

\end{tabular}
\caption{\label{tab:grouped} Rethermalization of $500$ inversely renormalized configurations with the Brower-Tamayo algorithm (IRG-BT) and separately with the Metropolis algorithm (IRG-M), and a comparison against randomly initialized configurations evolved with the Brower-Tamayo algorithm (Rand-BT). $20$ Metropolis sweeps suffice to remove the discrepancy in the nearest-neighbor operator $\langle nn\rangle$. The magnetization $\langle |m|\rangle$ is in agreement without rethermalization. }
\end{table}

We first implement constrained Monte Carlo methods to optimize the renormalization group transformation. As suggested in Ref.~\citep{PhysRevB.110.L140202}, we initiated our optimization of the renormalization group transformation using forward blocking, in a manner identical to the proposal of Swendsen for the optimization of real-space transformations~\citep{PhysRevLett.52.2321}. However, we subsequently discovered that the optimization of a transformation can be improved~\citep{Note1} if one optimizes directly on the space of the inversely renormalized configurations. Specifically, for an observable $O$ of the fine system and a configuration with coarse degrees of freedom $\Phi$ one calculates
\begin{equation}
\frac{\partial\avg{O}_{p(\phi | \Phi)}}{\partial\theta_a}
={\mathrm{Cov}_{p(\phi | \Phi)}\!\Big(O,\ \frac{\partial\ln p_\theta(\Phi|\phi)}{\partial\theta_a}\Big)},
\label{eq:lr}
\end{equation}
where $\theta_a$, $a=1,\dots,13$, are the variational parameters of the transformation $\theta=(K_1,\dots,K_{10},\alpha,c,\sigma)$. The covariance is
evaluated in the conditional distribution $p(\phi | \Phi)$ of each input configuration $\Phi$ and then averaged over the
input configurations. As an example, the observables most significantly improved by optimizing directly on the space of inversely renormalized configurations were the correlation functions $G_{200}$, $G_{210}$, $G_{300}$ and the plaquette $P$. By optimizing the transformation we approximately map the action onto itself, to allow for computational stability in the inversion of the renormalization group.

\paragraph{\label{sec:level6}Inverse renormalization group of 3D $\phi^{4}$ theory  with constrained Monte Carlo.---} 

To implement constrained Monte Carlo for the inversion of the renormalization group transformation, we sample with the Metropolis algorithm $3.6 \times 10^{3}$ uncorrelated configurations for lattice volume $V^{(0)}=16^{3}$ and the numerically calculated critical couplings $\lambda=1.1$, $2\kappa_c=0.3750966$.  We then initialize a fine system~\citep{Note1} with lattice volume $V^{(1)}=32^{3}$ and sample the degrees of freedom of the inversely renormalized system so that they remain consistent with the ones of the coarse system using constrained Monte Carlo with $10^{2}$ sweeps. We repeat this process twice, using as input the $V^{(1)}=32^{3}$ configurations and then the $V^{(2)}=64^{3}$ until we obtain a system of volume $V^{(3)}=128^{3}$. A configuration obtained from the inversion of the renormalization group via constrained Monte Carlo is depicted in Fig.~\ref{fig:fig1}.  

We define one sweep in the inversion of the renormalization group  as the update of $V$ selected lattice sites equal to the volume of the fine system. In Fig.~\ref{fig:fig2} we depict the equilibration of the constrained Monte Carlo for  all considered lattice volumes using as operators the nearest-neighbor interaction and the absolute value of the magnetization.  We observe that only a moderately small amount of sweeps is required to equilibrate the lattice which is practically constant for different lattice volumes. The dynamic critical exponent is calculated as $z=0.03(2)$ using the integrated autocorrelation times~\citep{Note1}.  We clarify that the depicted equilibration is that of $p(\phi | \Phi)$, not a rethermalization based on $p(\phi)$.

In Fig.~\ref{fig:fig3} we show the autocorrelation of the configurations using the nearest-neighbor operator for the case of the originally sampled uncorrelated configurations at lattice volume $V^{(0)}=16^{3}$ and the inversely renormalized configurations of $V^{(3)}=128^{3}$ obtained via constrained Monte Carlo. The integrated autocorrelation time $t_{\rm int}$ for $L=16$ and $L=128$ is $0.49(4)$ and $0.50(4)$, respectively, with the configurations remaining uncorrelated after multiple inversions of the renormalization group. 

In Fig.~\ref{fig:fig4} we rethermalize the inversely renormalized configurations with the Brower-Tamayo algorithm~\citep{PhysRevLett.62.1087} with Swendsen-Wang cluster updates~\citep{PhysRevLett.58.86}. We compare against the thermalization of random chains  at the same lattice volume using the Brower-Tamayo algorithm and, separately, the Metropolis algorithm. We observe that the uncorrelated configurations obtained at $V^{(3)}=128^{3}$ via three inversions of the renormalization group can be considered as near-equilibrium configurations. 

In Table~\ref{tab:grouped} we provide the expectation values of the nearest-neighbor operator and the absolute value of the magnetization. These correspond to the directly obtained inversely renormalized configurations at $0$ sweeps, and their rethermalization at $20$ and $500$ sweeps with the Brower-Tamayo algorithm and, separately, the Metropolis algorithm. We additionally list for reference the expectation values for randomly initialized configurations that are evolved with the Brower-Tamayo algorithm. We observe that the absolute value of the magnetization is within agreement without rethermalization, and that $20$ Metropolis sweeps suffice to remove the discrepancy from the nearest-neighbor operator.

The configurations constructed from $V^{(0)}=16^{3}$ to $V^{(3)}=128^{3}$ have all been proposed in the constrained Monte Carlo according to the action $S$ whose couplings $\lambda$ and $\kappa$ correspond to a numerical calculation of the critical point, and which is an approximation to the fixed point action $S \approx S'$. We retain only the $V^{(3)}=128^{3}$ configurations, with Metropolis rethermalization of $20$ sweeps, and we discard all other configurations. We then apply the forward renormalization group iteratively until we construct $3.6 \times 10^{3}$  configurations of $V=8^{3}$. By applying the forward renormalization group we allow the configurations to flow freely towards the fixed point action $S'$ of the transformation. We consider a set of $13$ even operators and $11$ odd operators~\citep{Note1} to calculate the critical exponents with the linearized renormalization group transformation matrix~\citep{PhysRevLett.42.859}. The results for $y_{h}$ and $y_{t}$ are depicted in Table~\ref{tab:1}, and at $16 \to 8$ they are in agreement with the 3D Ising universality class.

\begin{table}[t!]
\centering
\begin{tabular}{|l|c|c|c|c|}\hline
       & $128\to 64$ & $64\to 32$ & $32\to 16$ & $16\to 8$\\\hline
 $y_t$ & $1.593(8)$ & $1.593(7)$ & $ 1.599(9)$ & $1.596(10)$ \\\hline
$y_h$  & $2.4793(7)$ & $2.4834(4)$ & $2.4826(5)$ & $2.4822(12)$\\\hline
\end{tabular}
\caption{\label{tab:1}Calculations of the $y_{h}$ and $y_{t}$ exponents of the 3D $\phi^{4}$ quantum field theory with the linearized renormalization group transformation matrix using $13$ even and $11$ odd operators with $20$ Metropolis rethermalization sweeps. }
\end{table}

\paragraph{\label{sec:level7}Conclusions.---} 

In this Letter we introduced constrained Monte Carlo simulations in the inversion of renormalization group transformations within quantum field theory. Specifically, we optimized a transformation directly on the inversely sampled configurations to establish computational stability for high-dimensional systems. It removes the necessity to simulate a complicated fixed point action with multiple terms as observed before in the literature~\citep{PhysRevLett.89.275701}.  The method is further substantiated by a proof of exactness for the inverse renormalization group in the case of the fixed point and a near-exact inversion in the vicinity of the fixed point. Starting from a set of uncorrelated configurations at lattice volume $V=16^{3}$ for the three-dimensional $\phi^{4}$ theory we have iteratively inverted the renormalization group to construct without critical slowing down a set of uncorrelated and near-equilibrium configurations for lattice volume $V=128^{3}$  and we have extracted two critical exponents. 
 
 Inverse renormalization group methods with constrained Monte Carlo allow for an efficient generation of coarse-to-fine configurations, they are interpretable as they are established on insights of mathematical physics, and they are not subject to the significant algorithmic complexity of neural network architectures. They might therefore provide a compelling setting for numerical calculations at the frontier of experimental science for which there exist substantial computational demands.

\paragraph{\label{sec:level8}Data Availability Statement.---} Data which support the results of this manuscript are available from the author upon request.  The values of the transformation parameters are provided in the Supplemental Material~\citep{Note1}. 

\paragraph{\label{sec:level9}Acknowledgements.---}   The author acknowledges support from the ``Simons Collaboration on the Physics of Learning and Neural Computation". Part of this research was conducted when the author was affiliated with Queen Mary University of London and the author acknowledges support from the Science and Technology Facilities
Council (STFC) Consolidated Grant ST/X00063X/1 ``Amplitudes, Strings \& Duality".

\bibliography{ms}
\newpage
\onecolumngrid  

\thispagestyle{empty}
\noindent
\begin{minipage}[t][\paperheight]{\paperwidth}
  \vspace*{-1in}%
  \hspace*{-0.8in}%
  \includegraphics[page=1, width=\paperwidth, height=\paperheight, keepaspectratio]{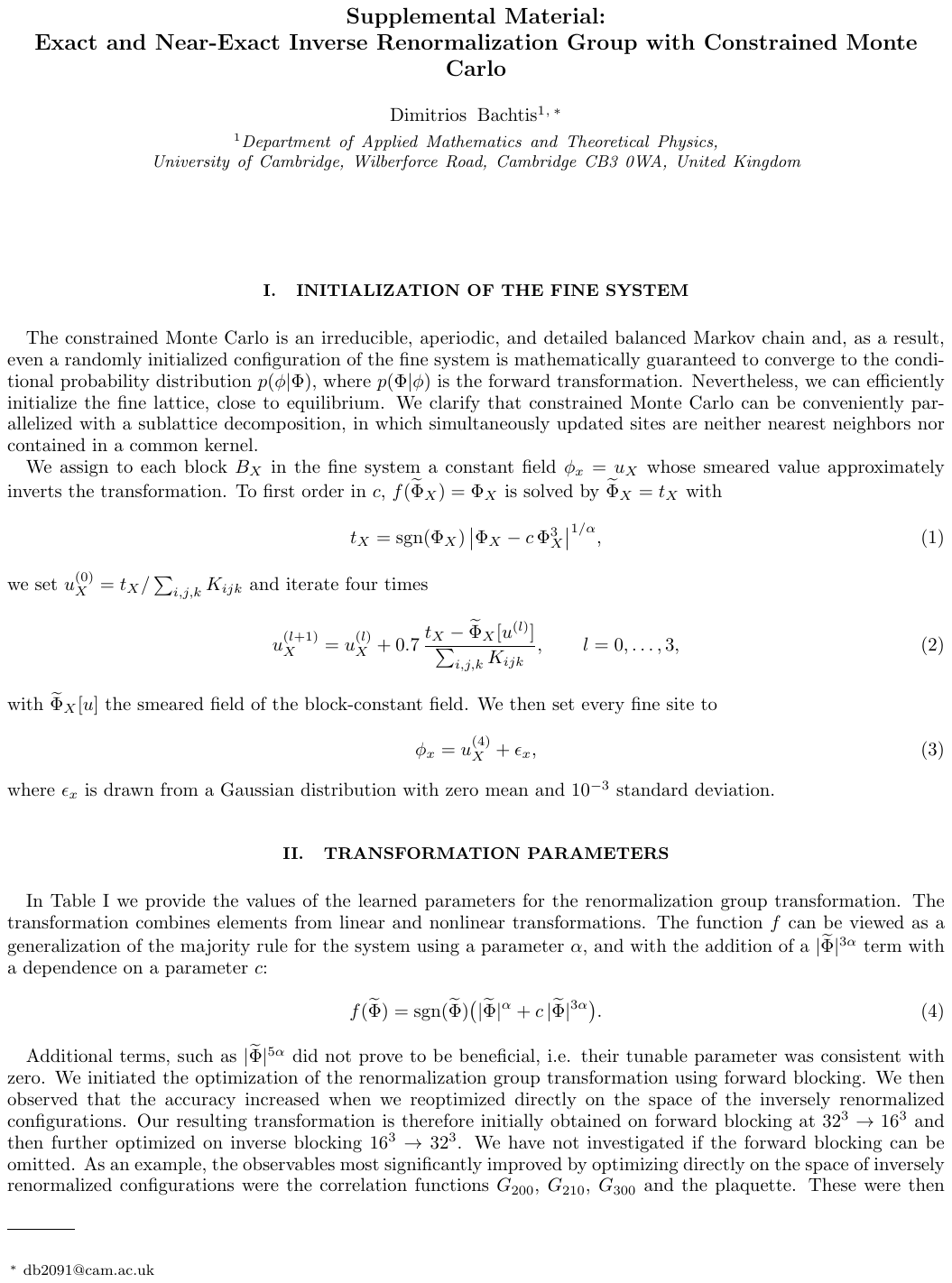}%
\end{minipage}
\clearpage

\thispagestyle{empty}
\noindent
\begin{minipage}[t][\paperheight]{\paperwidth}
  \vspace*{-1in}%
  \hspace*{-0.8in}%
  \includegraphics[page=2, width=\paperwidth, height=\paperheight, keepaspectratio]{supl.pdf}%
\end{minipage}
\clearpage

\thispagestyle{empty}
\noindent
\begin{minipage}[t][\paperheight]{\paperwidth}
  \vspace*{-1in}%
  \hspace*{-0.8in}%
  \includegraphics[page=3, width=\paperwidth, height=\paperheight, keepaspectratio]{supl.pdf}%
\end{minipage}
\clearpage
\end{document}